\documentclass[runningheads]{llncs}
\usepackage[T1]{fontenc}
\usepackage{graphicx,verbatim}
\usepackage{lipsum}
\usepackage{array}
\usepackage{amsmath}
\usepackage{enumitem}
\usepackage{amsfonts}
\usepackage{graphicx}
\usepackage{amsfonts}
\usepackage{graphicx}
\usepackage{epsfig}
\usepackage{graphicx}
\usepackage{amsmath}
\usepackage{amssymb}
\usepackage{mathtools}
\usepackage{resizegather}
\usepackage{marvosym}
\usepackage{dsfont}
\usepackage{enumitem}
\usepackage{caption}
\usepackage{multirow}
\usepackage{hyperref} 
\usepackage{hyperref} % 必须加载以支持 \href
\usepackage{xurl}     % 可选：允许 URL 换行
\begin{document}
\title{Speech Audio Generation from Dynamic MRI via a Knowledge Enhanced Conditional Variational Autoencoder}
%
 %% Removed for anonymized MICCAI 2025 submission
\author{Yaxuan Li\inst{1}$^{\star}$ \and
Han Jiang\inst{2}\thanks{Equal contribution. $^{\dag}$Corresponding author.} \and
Yifei Ma\inst{1} \and
Shihua Qin\inst{3} \and
Jonghye Woo\inst{4} \and
Fangxu Xing\inst{4}$^{\dag}$ 
}
\authorrunning{Y. Li et al.}
\institute{Department of Computer Science, The University of Hong Kong, Pokfulam, Hong
Kong\and
School of Software Engineering, Xi'an Jiaotong University, Xi'an, Shaanxi, China
 \and
Wake Forest University School of Medicine, Medical Center Boulevard, Winston-Salem, NC, USA
\and Department of Radiology, Harvard Medical School, Boston 08544, USA \\
\email{fxing1@mgh.harvard.edu}
}
% \institute{Department of Radiology, Harvard Medical School, Boston 08544, USA \and
% School of Software Engineering, Xi'an Jiaotong University, Xi'an, Shaanxi, China
%  \and
% State Key Laboratory for Novel Software Technology, Nanjing University, China \and
% Wake Forest University School of Medicine, Medical Center Boulevard, Winston-Salem, NC, USA
% }

% \author{Anonymized Authors}  %% Added for anonymized MICCAI 2025 submission
% \authorrunning{Anonymized Author et al.}
% \institute{Anonymized Affiliations \\
%     \email{email@anonymized.com}}
\titlerunning{Speech Audio Generation from Dynamic MRI via a KE-CVAE}

\maketitle              % typeset the header of the contribution
\begin{abstract}
Dynamic Magnetic Resonance Imaging (MRI) of the vocal tract has become an increasingly adopted imaging modality for speech motor studies. Beyond image signals, systematic data loss, noise pollution, and audio file corruption can occur due to the unpredictability of the MRI acquisition environment. In such cases, generating audio from images is critical for data recovery in both clinical and research applications. However, this remains challenging due to hardware constraints, acoustic interference, and data corruption. Existing solutions, such as denoising and multi-stage synthesis methods, face limitations in audio fidelity and generalizability. To address these challenges, we propose a Knowledge Enhanced Conditional Variational Autoencoder (KE-CVAE), a novel two-step "knowledge enhancement + variational inference" framework for generating speech audio signals from cine dynamic MRI sequences. This approach introduces two key innovations: (1) integration of unlabeled MRI data for knowledge enhancement, and (2) a variational inference architecture to improve generative modeling capacity. To the best of our knowledge, this is one of the first attempts at synthesizing speech audio directly from dynamic MRI video sequences. The proposed method was trained and evaluated on an open-source dynamic vocal tract MRI dataset recorded during speech. Experimental results demonstrate its effectiveness in generating natural speech waveforms while addressing MRI-specific acoustic challenges, outperforming conventional deep learning-based synthesis approaches\addtocounter{footnote}{-4}\footnote{\url{https://github.com/YaxuanLi-cn/KE-CVAE}}. 

% \footnote{The model, python implementation, and generated data from this work will be made publicly available upon publication to support further research in the field.}

% Authors must provide keywords and are not allowed to remove this Keyword section.

\end{abstract}

\section{Introduction} 

Capturing real-time deformations of the vocal tract during human speech through medical imaging is essential for various speech research applications~\cite{toutios2016illustrating,lim20193d}, as accurately characterizing the functional behavior of vocal articulators has multiple clinical implications. Several imaging modalities are currently used for this task, including electromagnetic articulography (EMA), ultrasound, and magnetic resonance imaging (MRI). However, EMA requires attaching multiple sensors to the articulators, which can potentially disrupt natural speech patterns, and it can only track a limited number of points on the tongue's surface~\cite{perkell1992electromagnetic}. Although ultrasound is non-invasive, it has a restricted field of view and struggles to clearly visualize deeper structures such as the palate and pharyngeal walls~\cite{lulich2018acquiring}. In contrast, MRI offers a more effective solution by providing high-contrast anatomical imaging with high resolution while remaining non-invasive. Advances in MRI technology have led to the development of high-speed dynamic cine imaging, enabling rapid and real-time imaging for in vivo speech~\cite{lim2021multispeaker,jin2024optimization}.
% However, simultaneously recording speech audio during image acquisition remains challenging due to several factors: 1) Hardware constraints: Specialized MRI-compatible microphones are required to withstand the strong magnetic fields without interfering with either the MRI scanner or the image quality \cite{isaieva2021multimodal,jacobs2022image}. 2) Acoustic interference: MRI scanners produce substantial and disruptive noise levels. The unpredictability of this background noise, which varies depending on scanning sequences and parameters, complicates the task of capturing clear audio recordings \cite{ravicz2000acoustic,bresch2006synchronized}. 3) Data integrity issues: The quality of dynamic image sequence reconstructions can significantly impact the integrity of sound recordings. Synchronization between image and audio is difficult, and noise, missing data, or artifacts during MRI scans may lead to corrupted or incomplete audio segments \cite{catana2015motion,menchon2019reconstruction}. Despite these challenges, capturing speech waveforms remains a critical requirement in many speech imaging studies \cite{liu2023speech,zheng2024incorporating}.

However, simultaneously recording the subjects' speech audio signals during image acquisition remains challenging due to several factors. (1) Hardware constraints require specialized MRI-compatible microphones that function within strong magnetic fields without interference from the MRI scanner or to the image quality~\cite{jacobs2022image}. (2) Acoustic interference from MRI scanners (the loud pulse sequence sounds) generates substantial and unpredictable noise, varying with scanning protocols/parameters and making it difficult to capture clean recordings of human voice~\cite{bresch2006synchronized}. (3) Data integrity issues arise as the process of dynamic MRI reconstruction can affect sound recordings, with synchronization difficulties, MRI-induced noise, missing data or artifacts leading to corrupted or incomplete audio segments~\cite{menchon2019reconstruction}. Despite these challenges, capturing speech waveforms remains a critical requirement in many speech imaging studies~\cite{liu2023speech,zheng2024incorporating}.

Multiple solutions have been proposed to obtain high-quality speech sound files in MRI environments. Denoising techniques enhance audio clarity but can only reduce noise rather than fully eliminate it and cannot reconstruct missing speech segments~\cite{erturk2013denoising,jin2023enhancing}. More recently, an innovative approach has emerged that synthesizes speech directly from MRI data. Liu et al. \cite{liu2023speech} developed a system that converts 4D tagged-MRI data into audio using Non-negative Matrix Factorization (NMF). While effective for processing short phrases, this method relies on a precomputed sequence of deformation fields to synthesize audio, limiting its ability to directly process dynamic cine images. Additionally, its multi-step process is computationally intensive and may not generalize well to longer or more varied speech patterns. Although these methods represent significant progress, they still fall short of producing fully accurate and complete speech sound files for comprehensive research applications.

\begin{figure}[t]
\begin{center}
\includegraphics[width=1\linewidth]{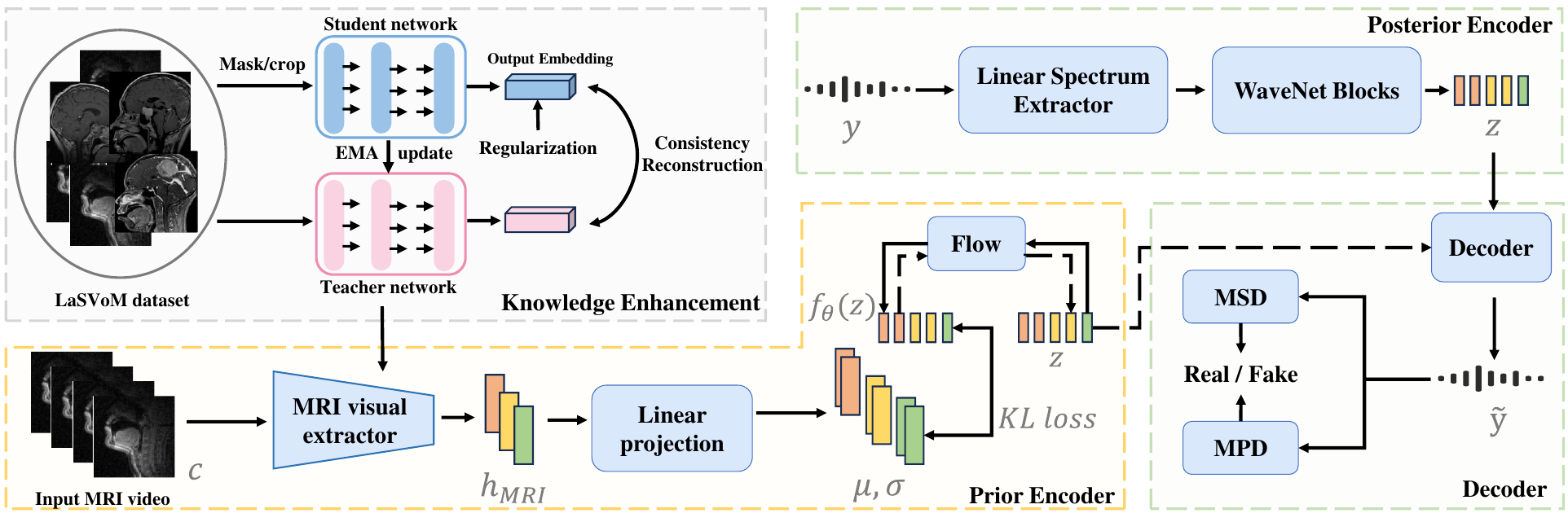}    
\end{center} 
\caption{Illustration of our KE-CVAE model. The dashed lines in the figure represent the inference process.} 
\label{fig:illus} 
\end{figure}

In non-medical imaging settings, several methods have been proposed for audio generation~\cite{kim2021conditional,lee2022hierspeech} that utilize an end-to-end variational inference framework, producing more natural-sounding signals than earlier two-stage models. Inspired by these approaches, we propose a Knowledge Enhanced Conditional Variational Autoencoder (KE-CVAE) with two key innovations: (1) A novel two-step “knowledge enhancement + variational inference” framework for speech audio generation from dynamic MRI sequences. Specifically, we have compiled a collection of thirteen public head and neck MRI image/video datasets for the training process using a self-supervised knowledge enhancement strategy. (2) Adoption of a variational inference framework to enhance the expressive power of generative modeling (Figure~\ref{fig:illus}). We conducted comprehensive experiments to generate speech sound using the open-access speech dynamic MRI dataset described in \cite{lim2021multispeaker}, demonstrating the effectiveness of KE-CVAE over conventional CNN and transformers in multiple metrics such as correlation, PESQ, and a subjective MOS score (detailed in the results section).
 
\section{Methods}
 
\subsection{Knowledge Enhancement with Domain-Specific Data}
\label{sec:knowledge enhancement}

Recent advances in self-supervised pre-training on images \cite{chen2021exploring,oquab2023dinov2} in computer vision have demonstrated that it is possible to learn robust and meaningful low-dimensional features without explicitly labeled supervision. Inspired by this concept, we propose a self-supervised knowledge enhancement strategy to learn latent variables from large-scale MRI data. Specifically, our model comprises a teacher network and a student network (Figure~\ref{fig:illus}), both based on vision transformers (ViTs)~\cite{liu2021swin}, denoted as $F_{{teacher}}$ and $F_{{student}}$, respectively. By leveraging a large-scale curated set of vocal tract MRI, our strategy enables the visual extractor to focus on articulatory-relevant regions and capture subtle distinctions in vocal tract configurations during speech production. To effectively learn domain-specific knowledge without annotations, we introduce three complementary loss functions: consistency loss for representation alignment, reconstruction loss for masked image modeling, and KoLeo regularization for feature distribution optimization.

\noindent \textbf{Consistency Loss. }The consistency loss is designed to maximize the alignment of the output classification ($[CLS]$) token embeddings between $F_{{teacher}}$ and $F_{{student}}$, facilitating better representation learning. Specifically, we pass the $[CLS]$ token through a standard multilayer perceptron (MLP) model to generate a vector of scores, followed by a softmax function to obtain the pseudo-class probability $p$. We compute consistency loss at both the global and local levels, corresponding to image-level and patch-level augmentations, respectively. The global consistency loss is defined as the cross-entropy loss between the teacher output $p_t$ and the student output $p_s$. The local consistency loss is computed as the cross-entropy loss between the $[CLS]$ embeddings of both networks, $p_{sl}$ and $p_{tl}$. We apply Sinkhorn-Knopp centering~\cite{caron2020unsupervised} to the teacher network output to enhance distribution alignment. The final consistency loss integrates both global and local components, formulated as follows:
% \begin{equation}
% \mathcal{L}_{con} = - \sum p_t \log p_s - \sum p_{ti} \log p_{si}   
% \end{equation}
\begin{equation}
\mathcal{L}_{con} = - \sum_{i=1}^{N} p_t^{(i)} \log p_s^{(i)} - \sum_{i=1}^{N} \sum_{j=1}^{M} p_{tl}^{(i,j)} \log p_{sl}^{(i,j)},
\end{equation}
where $N$ is the batch size and $M$ is the number of patches per sample.

\noindent \textbf{Reconstruction Loss. }
% Masked image modeling \cite{bao2021beit,zhou2021ibot} has been proposed in several recent works. 
We randomly replace the patches in the input image with masks using the $[MASK]$ token or random patch features with a certain probability. The masked input MRI image $\mathbf{V}_i$ can be represented as $\mathbf{V}_{i\_{mask}}$ where the embeddings of some patches $\{v_i\}_{i \in B}$ in $\mathbf{V}_{i\_{mask}}$ are replaced by trainable $[MASK]$ token embeddings. The reconstruction loss is computed by comparing the output embeddings of the $[MASK]$ tokens from both the teacher and student models:
\begin{equation}
\mathcal{L}_{{rec}} = - \sum_{i=1}^{N} \sum_{j=1}^{M} h_j \cdot F_{teacher}(\mathbf{V}_i) \log F_{student}({\mathbf{V}_{i\_mask}}),
\end{equation}
where $h_j = 1$ indicates that the token at position $j$ has been masked, and $h_j = 0$ indicates otherwise. Incorporating this reconstruction loss complements the consistency loss, facilitating the learning of robust MRI image embeddings at multiple levels.

\noindent \textbf{KoLeo Regularization Loss. }The KoLeo regularization loss~\cite{delattre2017kozachenko} has been designed to encourage a uniform span of the features within a batch. Given a batch of $ N $ output $[CLS]$ token embeddings from $\{F_{{student}}(\mathbf{V}_i)[CLS]\}_{i=1}^{N}$ of the student network, KoLeo regularization loss is defined as:
\begin{equation}
\mathcal{L}_{KoLeo} = -\frac{1}{N} \sum_{i=1}^{N} \log(d_{N,i}),
\end{equation}
where $ d_{N,i} = \min_{j \neq i} \| F_{{student}}(\mathbf{V}_i)[CLS] - F_{{student}}(\mathbf{V}_j)[CLS] \| $ is the minimal distance between $ F_{{student}}(\mathbf{V}_i)[CLS] $ and any other student network $[CLS]$ embedding within the batch. $ F_{{student}}(\mathbf{V}_i)[CLS] $ is also $\ell_2$-normalized before computing the KoLeo regularization loss.

\subsection{Variational Inference Framework on Dynamic Speech MRI}

We formulate the proposed model KE-CVAE as a conditional variational autoencoder. In this framework, $p(y|z)$ represents the likelihood function for generating an audio waveform data point $y$ given the latent variable $z$, while $q(z|y)$ denotes the approximate posterior distribution that infers $z$ from $y$. Additionally, $p(z|c)$ describes the prior distribution of the latent variables $z$, conditioned on the MRI sequence $c$. The objective of the CVAE is to maximize the variational lower bound, also known as the evidence lower bound (ELBO), of the intractable marginal log-likelihood of the data $\log p(y|c)$. This objective can be decomposed into two terms: the reconstruction loss and the KL divergence, expressed as follows:

\begin{equation}
\mathcal{L}_{ELBO} = \mathcal{L}_{{recon}} + {KL}(q(z|y) \,||\, p(z|c)).
\end{equation}

The reconstruction loss, \(\mathcal{L}_{{recon}}\), is defined as \(\mathcal{L}_{{recon}} = \left\| y_{{mel}} - \tilde{y}_{{mel}} \right\|_{1}\), where \(y_{{mel}}\) denotes the mel-spectrogram of the input audio, and \(\tilde{y}_{{mel}}\) represents the mel-spectrogram reconstructed by the decoder. The \(\left\| \cdot \right\|_{1}\) notation refers to the \(\ell_1\) norm.

The KE-CVAE model primarily comprises three components: a posterior encoder, a prior encoder, and a decoder. We detail the three modules as follows:

\noindent \textbf{Posterior Encoder. }The posterior encoder extracts the latent representation $z\sim q(z|y)$ from the input waveform $y$. We first transform the raw waveform into its linear spectrum. Several non-causal WaveNet residual blocks \cite{kim2020glow} are applied to extract an embedding sequence. Then we employ a linear layer to project the mean and variance from the normal posterior distribution $p(z|y)$.

% \noindent \textbf{Posterior Encoder. }The posterior encoder extracts the latent representation $z\sim q(z|y)$ from the input waveform $y$. We firstly transform the raw waveform to the linear spectrum. Then several non-causal WaveNet residual blocks \cite{prenger2019waveglow,kim2020glow} are applied to extract an embedding sequence.  Finally, we employ a linear layer to project the mean and variance of the normal posterior distribution $p(z|y)$.

\noindent \textbf{Prior Encoder. }In our proposed setting, dynamic MRI sequences serve as the condition for speech generation. The prior encoder models the conditional prior distribution $p(z|c)$ given the MRI sequence $c$. 
% Given the condition of the input MRI sequence $c$, the prior encoder provides the prior distribution $p(z|c)$. 
As described in Section \ref{sec:knowledge enhancement}, the pre-trained MRI visual extractor $F_{{teacher}}$ is applied in this module to obtain the hidden representation $h_{MRI}$. The linear projection layer following the blocks produces the mean $\mu$ and variance $\sigma$ of the normal posterior distribution. To further improve the scalability of the approximated posterior distributions, we use a normalizing flow $f_{\theta}$ to apply a sequence of invertible transformations \cite{rezende2015variational}. Therefore, the KL divergence is calculated by:
\begin{equation}
\begin{gathered}
    {KL}(q(z|y) \,||\, p(z|c)) = \log q(z|y) - \log p(z|c), \\
    p(z|c) = \mathcal{N}(f_{\theta}(z); \mu(c), \sigma(c)) \left| \det \frac{\partial f_{\theta}(z)}{\partial z} \right|.
\end{gathered}
\end{equation}
where $\det$ denotes the determinants.

\noindent \textbf{Decoder. }The decoder uses the latent variable \( z \) to reconstruct the waveform $\tilde{y} \sim p(y|z)$. In addition, we adopt the adversarial training strategy. The discriminator $D$ follows HiFi-GAN’s multi-period discriminator (MPD) and multi-scale discriminator (MSD) architecture \cite{kong2020hifi}. Specifically, the adversarial losses \cite{mao2017least} for the generator $G$ and the discriminator $D$ are defined as:
\begin{equation}
\mathcal{L}_{adv}(D) = \mathbb{E}_{(y, z)} \left[ (D(y) - 1)^2 + (D(G(z)))^2 \right],    
\end{equation}
\begin{equation}
\mathcal{L}_{adv}(G) = \mathbb{E}_z \left[ (D(G(z)) - 1)^2 \right].  
\end{equation}
where $\mathbb{E}$ denotes the expectation. Specifically, we utilize the feature matching loss \cite{larsen2016autoencoding} as an element-wise reconstruction loss for more stable training procedure.

\section{Experiments and Results}
 \subsection{Datasets and Evaluation Metrics}
\textbf{LaSVoM Dataset.} To address the scarcity of large-scale head and neck MRI resources, we constructed a ``large-scale vocal track MRI'' data collection (LaSVoM) by aggregating and curating data from thirteen public MRI image and video datasets in the head and neck region. Through systematic frame extraction and resizing, we collected over 30,000 high-quality mid-sagittal head and neck MRI images, each resized to focus on the vocal tract region with a size of 84×84 pixels. These images encompass a wide range of phonetic contexts, speakers, and articulation patterns, providing a comprehensive resource for the proposed knowledge enhancement process. 

\noindent \textbf{Variational Inference Dataset.} We utilized the open dynamic MRI in speech dataset in \cite{lim2021multispeaker}, which consists of cine MRI capturing dynamic vocal tract movements in the sagittal plane, accompanied by synchronized audio recordings. The dataset includes cine sequences from 75 participants performing various speech tasks. For training and testing, we split the dataset into an 8:2 ratio.

\noindent \textbf{Evaluation Metrics.} We perform all experiments in a speaker-independent manner so that the metrics we used are not affected by the change of speakers. We calculated conventional objective metrics for audio waveform evaluation, including 2D Pearson’s correlation coefficient (Corr2D)~\cite{chi2005multiresolution} and perceptual evaluation of speech quality (PESQ)~\cite{recommendation2001perceptual}. We also performed a subjective MOS (mean opinion score) test to evaluate all compared methods. Specifically, we randomly selected 30 from 1662 test audios for subjective listening and asked 10 human raters to assess their quality. The raters evaluated the audio quality by comparing each reconstructed audio sample with its corresponding ground truth, rating them on a scale from 1 to 5 based on their similarity and overall quality.

 \subsection{Implementation and Training Protocols}

Due to significant imaging differences across scanner platforms, we normalized each image in the LaSVoM dataset to a mean of 0.5 and a standard deviation of 0.5. For both the training and test datasets, we applied a noise reduction algorithm\footnote{\url{https://github.com/timsainb/noisereduce}} to remove background noise from the audio, followed by audio magnitude normalization. Since the dataset contained a substantial number of silent segments (audio volume below -60 dB for more than 0.2 seconds), we used FFmpeg\footnote{\url{https://github.com/FFmpeg/FFmpeg}} to detect and segment audio sequences, retaining only the voiced segments. Finally, we resampled the audio to 16,000 Hz and the video to 80 fps.

In the proposed model, The MRI visual extractor was composed of 12 ViTs blocks. The dimension of the WaveNet was 512. The decoder and discriminator version of the HiFi-GAN we used was V1. The networks were trained using the AdamW optimizer \cite{loshchilov2017decoupled} with $\beta_1 = 0.8$, $\beta_2 =0.99$ and weight decay $\lambda = 0.01$. We implemented our approach using the Pytorch toolbox and trained for a total of 48 hours on an NVIDIA Tesla A100 GPU, where 40 epochs were used for the knowledge enhancement stage and 100 epochs were used for the variational inference stage.

\subsection{Benchmarking}

\begin{table}[t]
\centering 
\caption{Quantitative evaluation results of the audio quality using different synthesis methods on various metrics} 
\label{table:vanilla}  

\resizebox{0.85\linewidth}{!}{ 
\begin{tabular}{l|c|c|c|c}
\hline
% Method & \#Param & Corr2D for spectrogram $\uparrow$ & PESQ for waveform $\uparrow$ & MOS $\uparrow$ \\ 
Method & \#Param & Corr2D $\uparrow$ & PESQ  $\uparrow$ & MOS $\uparrow$ \\ 
\hline\hline
 
% Lip2AudSpect~\cite{akbari2018} w/o NMF & ?? & ?? & - & - \\ 
Vanilla (CNN) \textit{w/} KE & 124.0 M & 0.672 & 1.135 & 3.48 ($\pm$0.14) \\ 
Vanilla (CNN) & 124.0 M & 0.641 & 1.116 & 3.16 ($\pm$0.09) \\ 
Vanilla (Transformer) \textit{w/} KE & 146.5 M & 0.688 & 1.140 & 3.57 ($\pm$0.15) \\ 
Vanilla (Transformer) & 146.5 M & 0.649 & 1.131 & 3.22 ($\pm$0.12) \\ 
\hline
\textbf{Ours} & 121.1 M & \bf{0.818} & \bf{1.251} & \bf{4.13} ($\pm$\bf{0.08}) \\ 
\hline
Ours \textit{w/o} adversarial training & 60.0 M & 0.759 & 1.190 & 3.80 ($\pm$0.13) \\  
Ours \textit{w/o} Flow & 120.6 M & 0.798 & 1.237 & 4.02 ($\pm$0.10) \\ 
\hline
\end{tabular}}
\end{table}

\begin{table}[t]
\centering 
\caption{Ablation study in our knowledge enhancement strategy} 
\label{table:ablation}

\resizebox{0.65\linewidth}{!}{
\begin{tabular}{l|c|c|c}
\hline
% Method& Corr2D for spectrogram $\uparrow$ & PESQ for waveform $\uparrow$ & MOS $\uparrow$\\\hline\hline
Method& Corr2D $\uparrow$ & PESQ $\uparrow$ & MOS $\uparrow$\\\hline\hline

Baseline 	&0.716  &  1.152 & 3.74 ($\pm$ 0.12) \\\hline

$+\mathcal{L}_{con}$ &0.758  &  1.170 & 3.88 ($\pm$ 0.14) \\

$+\mathcal{L}_{con}+\mathcal{L}_{rec}$ &0.802  &  1.226 & 4.06 ($\pm$ 0.10) \\ 

$+\mathcal{L}_{con}+\mathcal{L}_{rec}+\mathcal{L}_{Koleo}$ &0.818  & 1.251 & 4.13 ($\pm$ 0.08) \\
\hline
\end{tabular}}
\label{table:ablation}  
\end{table}

\noindent \textbf{Quantitative Analysis.}
To validate the effectiveness of the proposed KE-CVAE framework, we compared it against two vanilla network variants following \cite{akbari2018lip2audspec,prajwal2020learning}. Specifically, we replaced the variational inference step with conventional CNN and transformer architectures with a similar number of parameters for the same speech audio generation task. In these vanilla variants, we directly used the MRI sequences as input and passed them through a series of encoders to obtain the audio output. Both vanilla models were optimized using the reconstruction loss $\mathcal{L}_{recon}$ as defined in our method. From Table~\ref{table:vanilla}, it is prominent that the performances of both architectures are evidently lower than that of KE-CVAE. The p-value obtained from the one-tailed Student's t-test is less than 0.05, indicating statistical significance. Notably, our knowledge enhancement step (KE in Table~\ref{table:vanilla}) also proves to be beneficial for the two vanilla methods, demonstrating its general applicability and effectiveness. 

We further conducted an ablation study to assess the contributions of different components in the KE-CVAE framework. Results in Table~\ref{table:vanilla} indicate that removing adversarial training and Flow mechanisms leads to a noticeable performance drop, highlighting the critical role of either component in achieving optimal performance. Meanwhile, every component in Section~\ref{sec:knowledge enhancement} plays an important role in the whole framework. To evaluate the impact of each objective function, we performed additional ablation studies and presented the results in Table~\ref{table:ablation}, where we used the CVAE model without knowledge enhancement as baseline. Results show that all the proposed loss functions in Section~\ref{sec:knowledge enhancement} are essential in improving the final performance.

\begin{figure}
    \centering
    \includegraphics[width=1\linewidth]{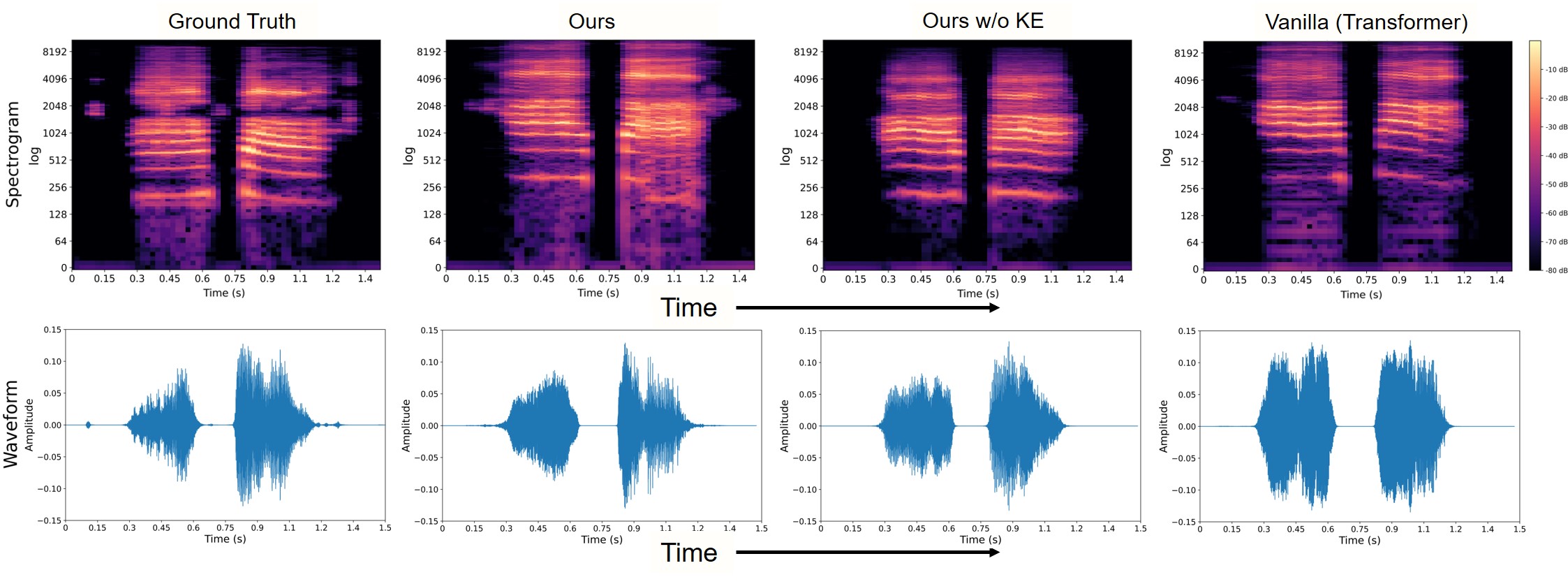}
    \caption{An example of generated audio spectrograms and waveforms using different methods compared to the ground truth.}
    \label{fig:visual}
\end{figure}

\noindent \textbf{Qualitative Analysis.}
Moreover, we visualized the spectrograms of the generated audio clips from the test stage, with one example shown in Figure~\ref{fig:visual}. As can be seen, the generated spectrogram and its corresponding audio waveform from KE-CVAE exhibit superior alignment with the ground truth. In contrast, the CVAE framework without the knowledge enhancement step (\textit{w/o} KE in Figure~\ref{fig:visual}) and the vanilla transformer model's quality are visually lower than that of the proposed full KE-CVAE.

\section{Conclusion}
In this work, we developed a novel two-step “knowledge enhancement + variational inference” framework to synthesize high-quality speech waveforms from dynamic MRI sequences, addressing critical challenges in synchronizing speech audio with real-time MRI recordings. In the knowledge enhancement phase, we proposed a robust self-supervised training pipeline that leverages large-scale MRI data to learn domain-specific features without explicit annotations. Following this, we integrated a variational inference framework to enhance the model's expressiveness. Specifically, we employed a posterior encoder, a prior encoder, and a decoder to effectively map latent variables to speech waveforms, further augmenting them with normalizing flow and adversarial training to improve the overall training process. Experimental results have demonstrated the efficacy of KE-CVAE in generating high-quality, temporally accurate speech from dynamic MRI data. This method has the potential to significantly enhance the accuracy of speech analysis in both clinical and research settings.

\subsubsection{Disclosure of Interests.} The authors have no competing interests to declare
that are relevant to the content of this article.

\bibliographystyle{splncs04}
\bibliography{Paper-2374}

\end{document}